\documentclass[conference]{IEEEtran}
\IEEEoverridecommandlockouts
\usepackage{cite}
\usepackage{amsmath,amssymb,amsfonts}
\usepackage{algorithmic}
\usepackage{graphicx}
\usepackage{textcomp}
\usepackage{hyperref}
\usepackage{xcolor}
\usepackage{pythontex} 
\def\BibTeX{{\rm B\kern-.05em{\sc i\kern-.025em b}\kern-.08em
    T\kern-.1667em\lower.7ex\hbox{E}\kern-.125emX}}
\begin{document}

\title{Wikidata-lite for \\
Knowledge Extraction and Exploration
}

\author{\IEEEauthorblockN{Phuc Nguyen}
\IEEEauthorblockA{\textit{National Institute of Informatics, Japan } \\
phucnt@nii.ac.jp}
\and
\IEEEauthorblockN{Hideaki Takeda}
\IEEEauthorblockA{\textit{National Institute of Informatics, Japan}\\
takeda@nii.ac.jp}

}

\maketitle

\begin{abstract}
Wikidata is the largest collaborative general knowledge graph supported by a worldwide community. It includes many helpful topics for knowledge exploration and data science applications. However, due to the enormous size of Wikidata, it is challenging to retrieve a large amount of data with millions of results, make complex queries requiring large aggregation operations, or access too many statement references. This paper introduces our preliminary works on Wikidata-lite, a toolkit to build a database offline for knowledge extraction and exploration, e.g., retrieving item information, statements, provenances, or searching entities by their keywords, attributes. Wikidata-lite has high performance and memory efficiency, much faster than the official Wikidata SPARQL endpoint for big queries. The Wikidata-lite repository is available at \url{https://github.com/phucty/wikidb}. 
\end{abstract}

\begin{IEEEkeywords}
Wikidata-lite, Wikidata, Knowledge Graphs
\end{IEEEkeywords}
\section{Introduction} \label{sec:intro}
Wikidata is the largest collaborative general knowledge graph supported by a worldwide community. As of November 2022, Wikidata contains 100 million items, 1.4 billion statements and covers many general topics for knowledge exploration and data science applications\footnote{Wikidata Statistics: \url{https://www.wikidata.org/wiki/Wikidata:Statistics}}. Wikidata also contains references that provide additional information, such as sources and URL links to such statements (1 billion statement references as in the dump November 2022 of Wikidata).

Unfortunately, due to the enormous size of Wikidata, extracting and exploring knowledge from Wikidata are challenging even using the standard public Wikidata Query Service (WDQS)\footnote{Wikidata Query: \url{https://query.wikidata.org/}} or parsing official released dumps\footnote{Wikidata Dumps: \url{https://dumps.wikimedia.org/wikidatawiki/entities/}} of Wikidata. 
The amount of information in Wikidata has been growing rapidly 39 times over seven years, from 3 GB (gzip compressed dump) in 2015 to 118 GB in 2022, and it keeps increasing as more and more data. Because of the execution timeout limit problem of Wikidata online query service \cite{wikidata}, users cannot make large queries with millions of results, execute complex queries requiring large aggregation operations, or access huge statement references. Moreover, hosting a local copy of Wikidata requires powerful servers (minimum hardware requirement: 16 CPU cores, 128GB RAM, and 1.5T SSD storage \cite{wikidata_limit}.)

This paper introduces our preliminary works on Wikidata-lite, a toolkit to help build a Wikidata database offline from the Wikidata dump to extract, explore, and analyze Wikidata knowledge \cite{wikidb}. Wikidata-lite aims to be fast and easy for Wikidata knowledge extraction and exploration for data scientists and application developers. With these objectives, we implement Wikidata-lite with key-value stores  (\textbf{lmdb}\footnote{lmdb: \url{http://www.lmdb.tech/doc/}, and python binding \cite{lmdb}}) as the underneath database. It allows fast reading and saves memory since the entire database is available in a memory map of a hard disk; there is no need to do extra copy data to memory. The tool also supports fully concurrent read access from multiple processes and threads.  

Wikidata-lite is an ongoing development project, the current release features are as follows:
\begin{itemize}
    \item Retrieving information about 100 million entities and 1.4 billion statements for knowledge exploration. The information includes item labels, descriptions, aliases in multilingual, and statement information. 
    \item Accessing more than 1 billion references for fact-checking applications.
    \item Searching entities by their attributes for subset item extraction. For example, it takes 0.02 seconds to get 5,868,897 males (wd:Q6581072) using Wikidata-lite while getting a timeout limit error using the online query service of Wikidata.
\end{itemize}

The rest of the paper is structured as follows. We describe the related work in Section \ref{sec:related_work}. We introduce the Wikidata-lite toolkit and the current representative use cases in Section \ref{sec:tool}. Finally, in Section \ref{sec:future}, we present conclusions and discuss the upcoming features of Wikidata-lite.  
\begin{figure*}[htbp]
\centering
\includegraphics[width=\textwidth]{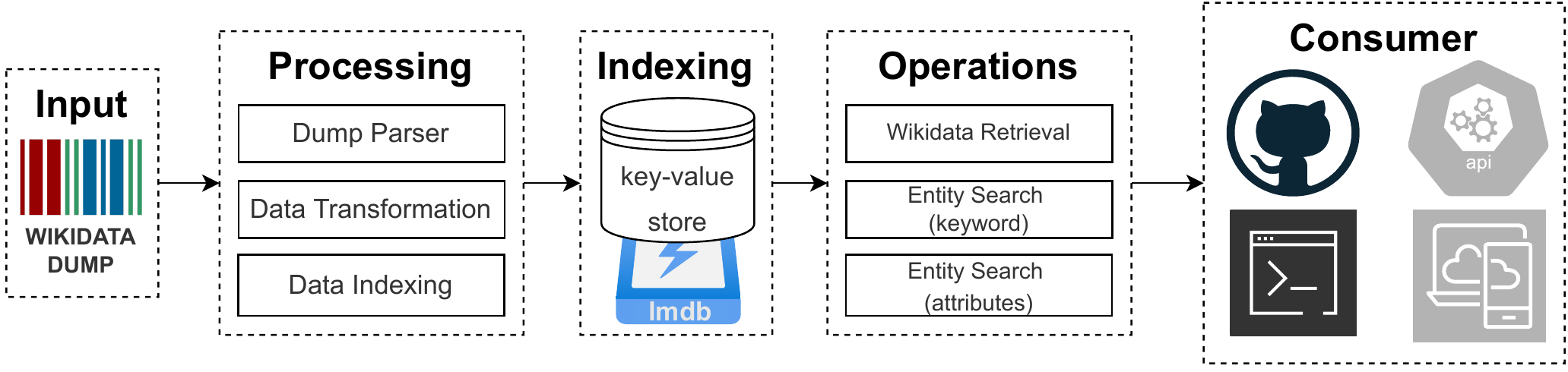}
\caption{Wikidata-lite workflow and features}
\label{fig:framework}
\end{figure*}
\section{Related Work} \label{sec:related_work}
Many tools are available to parse, query, and explore knowledge from Wikidata dumps. KGTK is a knowledge graph toolkit for creating, transforming, merging, and exploring knowledge graphs \cite{kgtk}. KGTK model knowledge graphs in tab-separated (TSV) files with four columns: edge-identifier, head, edge-label, and tail. Wikidata dumps could be converted to KGTK formats with transformation operations. KGTK provides many graph manipulation operations such as intersection, subtraction, and joining and other analytics operations such as calculating node popularities, degrees, shortest paths, or graph embeddings. The TSV files could be imported to the Kypher processor to query Wikidata knowledge \cite{kyper}. The Kyper underneath technique translates the TSV file into SQL and executes it on SQLite so that it also consumes an enormous memory if executing a large query. However, the current KGTK release does not support accessing statement references. Moreover, Wikidata-lite is implemented using lmdb on key-value storage with memory maps with \textbf{lmdb}, making it faster and simpler than using SQLite (one more abstraction of SQL data model on top of storage). 

WDumper is a toolkit for creating a topical subset (set of entities related to particular topics and the relationships between them) from the JSON dump of Wikidata \cite{wdumper}. Since WDumper allows users to extract and filter a small sampling of Wikidata, exploring subsets on small devices such as laptops or personal computers is feasible. The WDumper output is an RDF dump (N-Triple). Unlike the subset reusability objective of WDumper, Wikidata-lite focuses on extracting and exploring whole Wikidata knowledge.   

There are many other tools to query, transform and analyze RDF knowledge graphs, such as Neo4J\footnote{Neo4j: \url{https://neo4j.com/}}, Graphy\footnote{Graphy: \url{https://graphy.link/}} and RDFlib\footnote{RDFlib: \url{https://rdflib.readthedocs.io/en/stable/}}. In this work, we only focus on the JSON dump of Wikidata knowledge graph. 
\section{Wikidata-lite} \label{sec:tool}
In this section, we first present Wikidata-lite in Section \ref{sec:wikidata_lite}, then the use-cases are introduced in Section \ref{sec:usecases}.
\subsection{Architecture} \label{sec:wikidata_lite}

Wikidata-lite helps user access Wikidata offline with high-speed performance and memory efficiency usage. Figure \ref{fig:framework} depicts an overview of Wikidata-lite.

Given a JSON dump file of Wikidata, Wikidata-lite parses the dump as a JSON object for each item. Then, the item information is transformed into a key-value object, where keys are item IDs. The values are JSON objects about item labels, descriptions, aliases, site links, and all statements and references related to this item. In the data indexing module, the key-value objects are directly written to the memory map of a hard disk using \textbf{lmdb} library. The entire Wikidata is stored in memory maps allowing direct access to item information without copying this information to memory. We also build an inverted index of Wikidata items for item search with attribute features and calculate item popularity using the PageRank algorithms. The processing and indexing steps take two days, and the final \textbf{lmdb} file size is around 200GB (the compressed size for sharing is 35GB). 

The Wikidata retrieval module helps to fast access item information about the specific query as explained in Section \ref{sec:retrieval}. Users also can search items by labels (keywords) as mentioned in Section \ref{sec:search_keywords}. The item search by attributes allows users to find a subset of entities based on their attributes as introduced in Section \ref{sec:search_attributes}. We released the source code, a transformed database, and example codes in this preliminary work. The API and interfaces will be released soon in future work. 

\subsection{Use-cases} \label{sec:usecases}
This section explains Wikidata-lite use cases, i.e., item information retrieval, item search by keywords, and item search by attributes. 

\subsubsection{Item Information Retrieval} \label{sec:retrieval}
As all item information is stored in memory maps, it is really fast to retrieve items. The item information includes:
\begin{itemize}
\item Labels, descriptions, aliases: Users could get item labels, descriptions, and aliases in one specific language or all languages available. 
\item Site links: Users can access all Wikipedia titles and other projects linked to the item.
\item Statements and references: This function will return all statements and references of the query item. 
\item Others: Users also can retrieve item types, inverse relations such as the instance of relationships, the subclass of relationships, or getting relationships between two Wikidata items. 
\end{itemize}

The item information retrieval of Wikidata-lite is also easy to integrate into other downstream applications such as entity linking on text or tabular data \cite{semtab2021}, \cite{mtab2021}. 

\subsubsection{Item Search by Keywords} \label{sec:search_keywords}
We use the MTabES approach \cite{mtabes} to build item search by keywords. MTabES is a search tool that allows re-ranking between rankings of the algorithms BM25 (using ElasticSearch\footnote{ElasticSearch: \url{https://www.elastic.co/}}), fuzzy search with edit distances (using SymSpell:
Symmetric Delete algorithm \cite{Garbe_SymSpell_2012}), and PageRank scores. We index all item labels and aliases in multilingual. Users could access item search APIs and demos at \url{https://mtab.kgraph.jp/mtabes}.

\subsubsection{Item Search by Attributes} \label{sec:search_attributes}
We implement the item search by attributes inspired from the haswbstatement feature of Wikibase Cirrus Search\footnote{haswbstatement fetures: \url{https://www.mediawiki.org/wiki/Help:Extension:WikibaseCirrusSearch}}. The module returns items with specific values in the statement with specific properties. However, Wikidata-lite could directly search items with/or without specific properties as the inputs. 
For example, getting all scholarly articles that have relations with DNA, or X-ray diffraction, or molecular geometry, and Francis Crick and Nature. The corresponding query input is a combination of query statements of (AND, "P31", "Q13442814"), (OR, None, "Q7430"), (OR, None, "Q12101244"), (OR, None, "Q911331"), (AND, None, "Q123280"), (AND, None, "Q180445"). Wikidata-lite returns 46 items in 0.01754 sections for the query. A query statements contain three parts: Boolean operations (\textit{AND/OR/NOT}), property IDs (If there is no property given, the value of property ID is None), and item IDs. 

Another example of the queries that return millions of results is getting all male (AND, None, Q6581097). It takes 0.02 seconds to get 5,868,897 items, while the online query service of Wikidata will get a timeout limitation for this query. Other examples of item search by attributes can be found at \url{https://github.com/phucty/wikidb/blob/master/example.ipynb}. 

\section{Conclusion and Future Work} \label{sec:future}
This paper introduces our preliminary work on Wikidata-lite, a toolkit for offline extracting and exploring Wikidata knowledge. Wikidata-lite has high performance and memory efficiency, allowing users to retrieve item information, statements, and references from Wikidata dump. Users also can search items by keywords or attributes. 

One extension of Wikidata-lite is the ability to handle attribute item search with natural language text by integrating keyword item search to find correct item IDs beforehand. We also plan to use Wikidata-lite for data summarization tasks and natural language generation from Wikidata statements. 

On the consumer side, we plan to make Wikidata-lite available as a service released with REST APIs and an interface that allows users to use all Wikidata-lite operations directly. 
\section*{Acknowledgements}
The research was supported by the Cross-ministerial Strategic Innovation Promotion Program (SIP) Second Phase, “Big-data and AI-enabled Cyberspace Technologies” by the New Energy and Industrial Technology Development Organization (NEDO).

\bibliographystyle{IEEEtran}
\bibliography{IEEEabrv,ref.bib}


\end{document}